\RequirePackage{lineno}
\documentclass[aps,twocolumn,showpacs,preprintnumbers,amsmath,amssymb,superscriptaddress]{revtex4}%

\usepackage{graphicx}
\usepackage{dcolumn}
\usepackage{bm}
\usepackage{color}

\begin{document}

\preprint{preprint(\today)}

\title{Time-reversal invariant and fully gapped unconventional superconducting state in the bulk of the topological Nb$_ {0.25}$Bi$_{2}$Se$_{3}$}

\author{Debarchan Das}
\affiliation{Laboratory for Muon Spin Spectroscopy, Paul Scherrer Institute, CH-5232 Villigen PSI, Switzerland}

\author{K. Kobayashi}
\affiliation{Research Institute for Interdisciplinary Science, Okayama University, Okayama, 700-8530, Japan}

\author{M. P. Smylie}
\affiliation{Department of Physics and Astronomy, Hofstra University, Hempstead, New York 11549}

\author{C. Mielke III}
\affiliation{Laboratory for Muon Spin Spectroscopy, Paul Scherrer Institute, CH-5232 Villigen PSI, Switzerland}

\author{T. Takahashi }
\affiliation{Research Institute for Interdisciplinary Science, Okayama University, Okayama, 700-8530, Japan}

\author{K. Willa}
\affiliation{Institute for Solid State Physics, Karlsruhe Institute of Technology, Karlsruhe D-76021, Germany}

\author{J.-X.~Yin}
\affiliation{Laboratory for Topological Quantum Matter and Spectroscopy, Department of Physics, Princeton University, Princeton, New Jersey 08544, USA}

\author{U. Welp}
\affiliation{Materials Science Division, Argonne National Laboratory, 9700 S. Cass Ave., Lemont, Illinois 60439}

\author{M.Z.~Hasan}
\affiliation{Laboratory for Topological Quantum Matter and Spectroscopy, Department of Physics, Princeton University, Princeton, New Jersey 08544, USA}

\author{A.~Amato}
\affiliation{Laboratory for Muon Spin Spectroscopy, Paul Scherrer Institute, CH-5232
Villigen PSI, Switzerland}

\author{H.~Luetkens}
\affiliation{Laboratory for Muon Spin Spectroscopy, Paul Scherrer Institute, CH-5232
Villigen PSI, Switzerland}

\author{Z.~Guguchia}
\email{zurab.guguchia@psi.ch} \affiliation{Laboratory for Muon Spin Spectroscopy, Paul Scherrer Institute, CH-5232
Villigen PSI, Switzerland}

\begin{abstract}

Recently, the niobium (Nb)-doped topological insulator Bi$_2$Se$_3$, in which the finite magnetic moments of the Nb atoms are intercalated  in  the  van  der  Waals  gap between the Bi$_2$Se$_3$ layers, has been shown to exhibit both superconductivity with $T_{\rm c}$ ${\simeq}$ 3 K and topological surface states. Here we report on muon spin rotation experiments probing the temperature- and field-dependent of effective magnetic penetration depth $\lambda_{eff}\left(T\right)$ in the layered topological superconductor candidate Nb$_ {0.25}$Bi$_{2}$Se$_{3}$. The exponential temperature dependence of $\lambda_{eff}^{-2}$($T$) at low temperatures suggests a fully gapped superconducting state in the bulk with the superconducting transition temperature $T_{\rm c}$ = 2.9 K and the gap to $T_{\rm c}$ ratio 2${\Delta}$/$k_{\rm B}$$T_{\rm c}$ = 3.95(19). We also revealed that the ratio $T_{\rm c}$/$\lambda_{eff}^{-2}$ is comparable to those of unconventional superconductors, which hints at an unconventional pairing mechanism. Furthermore, time reversal symmetry breaking was excluded in the superconducting state with sensitive zero-field ${\mu}$SR experiments. We hope the present results will stimulate theoretical investigations to obtain a microscopic understanding of the relation between superconductivity and the topologically non-trivial electronic structure of Nb$_ {0.25}$Bi$_{2}$Se$_{3}$.

\end{abstract}


\maketitle

\section{Introduction}
Topological superfluidity, and superconductivity, are well-established phenomena in condensed matter systems. A very good example of a topological system is the A-phase of superfluid helium-3\cite{Levitin} which is a charge neutral topological superfluid. Topological superconductors (TSCs) are special families of those materials with unique electronic states, a full pairing gap in the bulk and gapless surface states consisting of Majorana fermions (MFs) \cite{Qi, Ando, Sato, Hasan,SXu}. Due to their scientific importance and potential applications in quantum computing, MFs have attracted lots of attention recently \cite{Kitaev,Wilczek,Beenakker}. This stimulating aspect of TSC motivated the condensed matter community to design novel topological systems by inducing superconductivity in a material with a non-trivial topological band structure.
Ways to realise TSC include through the proximity at the interface between
$s$-wave superconductors and topological insulators (TIs) with  large  spin-orbit coupling \cite{Beenakker}, at superconductor-magnet interfaces (1D chiral Majorana edge states) and at SC-TI-SC Josephson junctions (1D helical Majorana edge states). Another pathway to design TSC is to induce superconductivity in the bulk of TIs by doping, intercalation or pressure \cite{Novak, Sasaki, Matano, Zhong, Fu}. Currently, there are few potential material candidates for bulk TSC, obtained by following the latter approach: Sr$_{2}$RuO$_{4}$ \cite{LukeTRS}, which is generally believed to be a topological time reversal symmetry (TRS) breaking superconductor, the Weyl semimetal $T_{d}$-MoTe$_{2}$ \cite{GuguchiaMoTe2}, which is reported to be a TRS invariant topological superconductor, and Cu/Sr/Nb-doped Bi$_2$Se$_3$ created from one of the most studied TI Bi$_2$Se$_3$ \cite{Hor,Kriener,Krieger,Liu, Smylie3,Smylie1,Smylie2,Kobayashi}, which is reported to be a nematic \cite{FuPRB} superconductor.
The distinguishing feature of the Bi$_2$Se$_3$-derived superconductors is the emergence of a pronounced two-fold in-plane anisotropy of all superconducting properties \cite{Yonezawa2019} even though the crystal structure has three-fold symmetry (Fig. 1a).  This so-called nematic behavior was ascribed to an unconventional odd-parity two-component superconducting order parameter of E$_{u}$ symmetry which preserves time reversal symmetry \cite{FuPRB,Fu,Venderbos,Nagai}. The two-components, $\Delta_{\rm 4x}$ and $\Delta_{\rm 4y}$, are characterized by two in-plane point nodes and deep minima, respectively, defining the nematic axis.

\begin{figure*}[htb!]
\includegraphics[width=0.8\linewidth]{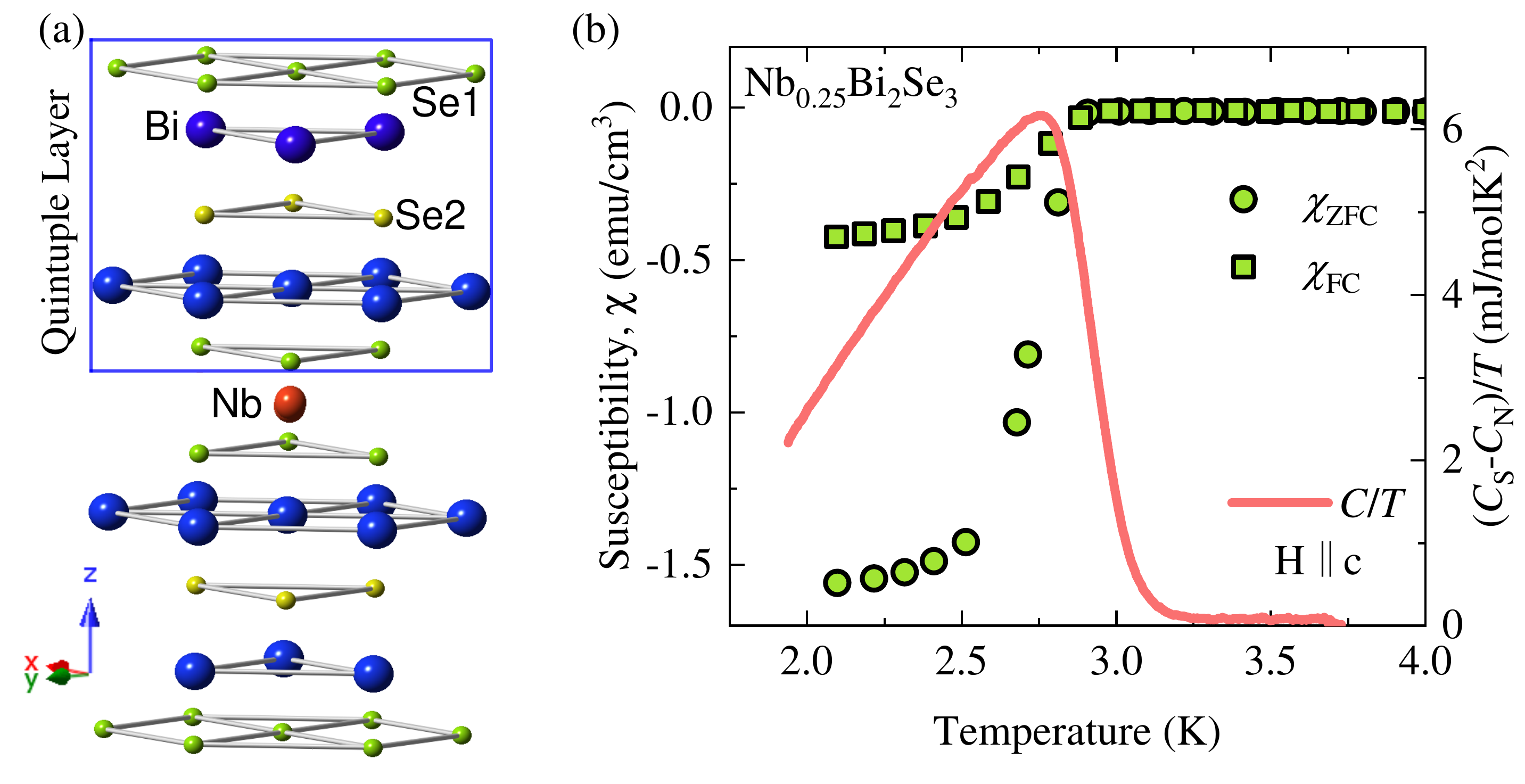}
\caption{(Color online) \textbf{Crystal structure and bulk superconducting transition for Nb$_ {0.25}$Bi$_{2}$Se$_{3}$.} (a) Crystal structure of Nb doped Bi$_2$Se$_3$ emphasizing the position of Nb intercalation between Bi$_2$Se$_3$ layers (highlighted as Quintuple layer). (b) Temperature dependence of dc magnetic susceptibility (left axis) in H($\parallel$$c$) = 1~mT and zero-field superconducting part of the specific heat divided by temperature (right axis) for the single crystalline sample of Nb$_ {0.25}$Bi$_{2}$Se$_{3}$. The susceptibility and specific heat measurements 
show similar transition temperature ($\sim$ 2.9~K).}
\label{fig7}
\end{figure*}

In this framework, the Nb-doped  Bi$_2$Se$_3$ system appears to be unique because macroscopic magnetic ordering below the superconducting critical temperature of 3.2 K has been reported\cite{Qiu} 
indicating a spontaneous spin rotation symmetry breaking of the Nb magnetic moments and TRS breaking. Nevertheless, detailed experimental studies involving, magnetoresistance, surface sensitive penetration depth measurements and Andreev reflection spectroscopy corroborate the existence of a nematic phase in Nb-doped Bi$_2$Se$_3$ system \cite{Smylie1,Smylie2, Asaba, Kurter, Shen}.  Due to the relatively large superconducting volume fraction in Nb-doped Bi$_2$Se$_3$ compared to its Cu-counterpart, it can be an ideal material to investigate the superconducting properties in detail enabling further insight into the physics of TSC. However, there are limited experimental studies reported exploring the bulk superconducting properties of this system. Thus, thorough exploration of superconductivity on the microscopic level in the bulk of Nb$_{0.25}$Bi$_2$Se$_3$ from both experimental and theoretical perspectives are required. In this regard, we concentrate on muon spin rotation/relaxation/resonance  ($\mu$SR) measurements of the magnetic penetration depth $\lambda$ in Nb$_{0.25}$Bi$_2$Se$_3$, which is one of the fundamental parameters of a superconductor, since it is related to the superfluid density $n_{s}$ via 1/${\lambda}^{2}$ = $\mu_{0}$$e^{2}$$n_{s}/m^{*}$ (where $m^{*}$ is the effective mass). Most importantly, the temperature dependence of ${\lambda}$ is particularly sensitive to the structure of the SC gap. Moreover, zero-field ${\mu}$SR is a very powerful tool for detecting a spontaneous magnetic field due to TRS breaking in exotic superconductors, because internal magnetic fields as small as 0.1 G are detected in measurements without applying external magnetic fields.

We report on fully gapped and a time reversal invariant superconducting state
in the bulk of topological system Nb$_{0.25}$Bi$_2$Se$_3$. We evaluated the higher limit of the zero-temperature penetration depth to be $\sim$ 930(10) nm. Interestingly, the $T_{\rm c}$/$\lambda_{eff}^{-2}$  ratio is comparable to those of unconventional superconductors. The relatively high $T_{\rm c}$ for small carrier density may hint at an unconventional pairing mechanism in Nb$_{0.25}$Bi$_2$Se$_3$.

\section{EXPERIMENTAL DETAILS }

 The details of the synthesis of the single crystal samples of Nb$_{0.25}$Bi$_{2}$Se$_{3}$ are reported elsewhere \cite{Kobayashi}. High purity samples were produced by properly selecting the crystals from the melt. Magnetization, X-ray diffraction, and specific heat experiments were performed on the single crystals, while   
 ${\mu}$SR experiments were measured on powdered crystals from the same batch.  X-ray diffraction analysis confirmed that the synthesized samples are single phase materials. The magnetization measurements were performed in a commercial SQUID magnetometer ($Quantum~Design$ MPMS-XL). AC specific heat was measured on a SiN membrane-based nanocalorimeter platform\cite{Willa}. The temperature dependence of the superconducting contribution \cite{Tagliati} (the normal state electronic specific heat is subtracted) was extracted by subtracting data with H($\parallel$$c$) = 1~T from the zero-field data \cite{WillaPRB}. The specific heat was scaled to match the published data\cite{Asaba} in the normal state as a precise determination of the sample volume was not possible due to the small mass of the sample.  
 
 Transverse-field (TF) and Zero-field (ZF) ${\mu}$SR experiments were performed at the ${\pi}$E1 beamline of the Paul Scherrer Institute (Villigen, Switzerland), using the Dolly spectrometer. The sample was pressed into a 5~mm pellet and mounted on a Cu holder using GE varnish. This holder assembly was then fixed to the $^3$He cryostat. The Spectrometer is equipped with a standard veto setup providing a low-background ${\mu}$SR signal. All TF experiments were carried out after a field-cooling procedure in the temperature range of 0.26 - 5~K. The ${\mu}$SR time spectra were analyzed using the free software package MUSRFIT \cite{Bastian}.

\begin{figure*}[htb!]
\includegraphics[width=0.9\linewidth]{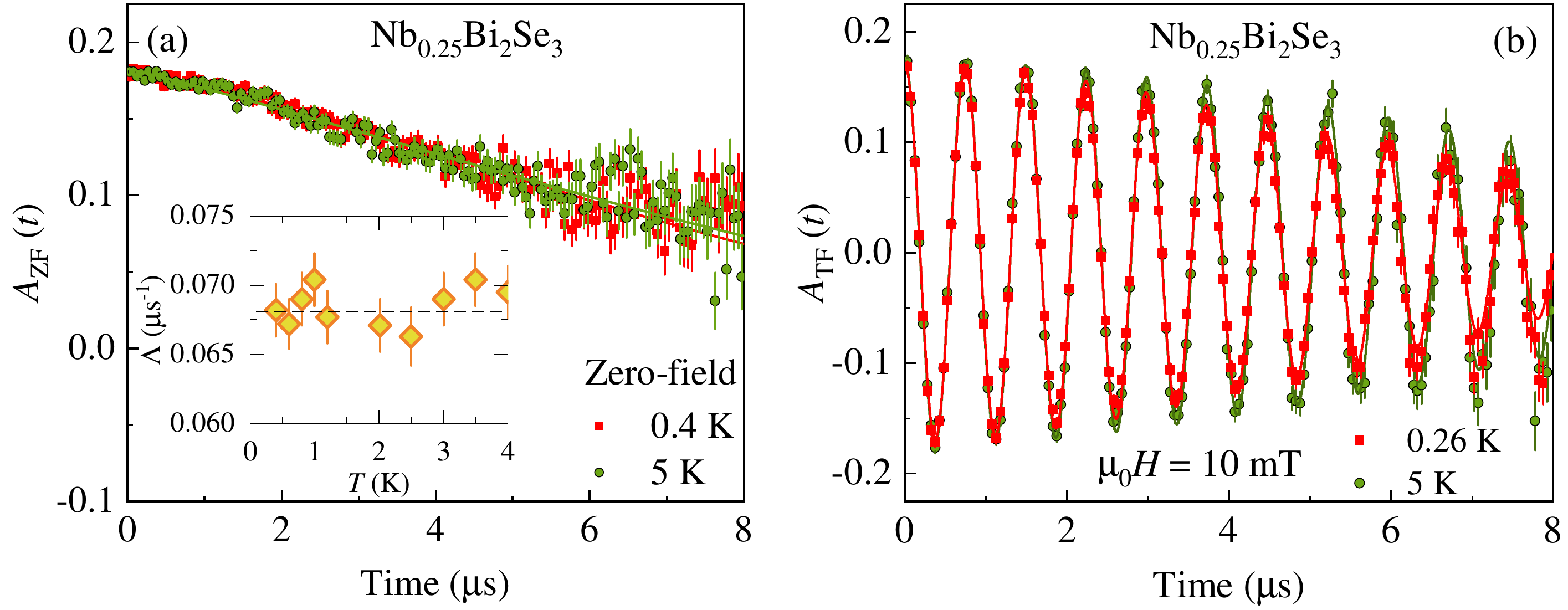}
\caption{(Color online) \textbf{Transverse-field (TF) and Zero-field (ZF) ${\mu}$SR time spectra for Nb$_ {0.25}$Bi$_{2}$Se$_{3}$.} Zero-field (ZF) ${\mu}$SR (a) and transverse-field (TF)(b) ${\mu}$SR time spectra obtained above and below $T_{\rm c}$ for Nb$_ {0.25}$Bi$_{2}$Se$_{3}$ (after field cooling the sample from above $T_{\rm c}$). Inset of (a) shows the temperature dependence of the electronic relaxation rate measured in zero magnetic field of Nb$_{0.25}$Bi$_2$Se$_3$ with $T_{\rm c}$ $\sim$ 3 K. Error bars are the s.e.m. in about 10$^{6}$ events. The error of each bin
count n is given by the s.d. of n. The errors of each bin in $A(t)$ are then calculated by s.e. propagation.}
\label{fig7}
\end{figure*}

\section{RESULTS}

Figure~1a shows the anisotropic layered rhombohedral crystal structure of Nb-intercalated Bi$_{2}$Se$_{3}$. Nb atoms are residing in van der Waals gap between adjacent Bi$_{2}$Se$_{3}$ quintuple Se-Bi-Se-Bi-Se layers (five covalently bonded atomic sheets) as shown in the figure. The quintuple layers are weakly bonded together by van der Waals interaction. The left axis of Fig~1b shows the temperature dependence of field cooled (FC) and zero field cooled (ZFC) magnetic susceptibility $\chi(T)$ for the single crystal in an applied magnetic field of H($\parallel$$c$) = 1 mT.  A clear diamagnetic transition appears in both FC and ZFC plots with a critical temperature $T_{\rm c}$ ${\simeq}$ 2.9 K and with 100 ${\%}$ superconducting volume fraction, confirming the onset of bulk superconductivity in studied Nb$_ {0.25}$Bi$_{2}$Se$_{3}$. To elucidate the intrinsic and bulk nature of this superconducting phase transition, we measured specific heat on a single crystalline sample of Nb$_ {0.25}$Bi$_{2}$Se$_{3}$. The right axis of the Fig~1b shows the temperature variation of the superconducting contribution to the specific heat divided by the temperature which exhibits a prominent step-like anomaly at $T_{\rm c}$, indicating bulk superconductivity in this system.

First, we have carried out ZF-${\mu}$SR experiments above and below $T_{{\rm c}}$ to search for possible magnetism (static or fluctuating) in Nb$_ {0.25}$Bi$_{2}$Se$_{3}$. As shown in Fig. 2a, no sign of either static or fluctuating magnetism could be detected in ZF time spectra down to 0.4 K. The ZF $\mu$SR spectra can be well described by a damped Gaussian Kubo-Toyabe (KT) depolarization  function \cite{Toyabe}, reflecting the field distribution at the muon site created by the nuclear moments. Moreover, no change in ZF-${\mu}$SR relaxation rate across $T_{c}$ was observed (see inset of Fig~2a), pointing to the absence of any spontaneous magnetic fields associated with a TRS \cite{LukeTRS,HillierTRS} breaking pairing state in Nb$_ {0.25}$Bi$_{2}$Se$_{3}$. This demonstrates that there is, in fact, no magnetic ordering in Nb$_ {0.25}$Bi$_{2}$Se$_{3}$.

Figure~2b represents the TF-$\mu$SR spectra for Nb$_ {0.25}$Bi$_{2}$Se$_{3}$, measured in an applied magnetic field of 10 mT above (5 K) and below (0.26 K) the SC transition temperature $T_{\rm c}$.
Above $T_{\rm c}$ the oscillations show a small relaxation due to the random local
fields from the nuclear magnetic moments. At 0.26 K, the relaxation rate increases due to the presence of a nonuniform local field distribution as a result of the formation of a flux-line lattice (FLL) in the SC state.
As indicated by solid lines  in Fig~2a, TF ${\mu}$SR data were analyzed by using the following functional form:\cite{Bastian}
\begin{equation}
A_{TF_S}(t)=A_Se^{\Big[-\frac{(\sigma_{sc}^2+\sigma_{nm}^2)t^2}{2}\Big]}\cos(\gamma_{\mu}B_{int}t+\varphi),
\label{eq1}
\end{equation}
Here $A_S$ denotes the initial asymmetry, $\gamma_\mu/(2{\pi})\simeq 135.5$~MHz/T is the muon gyromagnetic ratio, and ${\varphi}$ is the initial phase of the muon-spin ensemble. $B_{\rm int}$ represents the internal magnetic field at the muon site, the relaxation rates ${\sigma}_{\rm sc}$ and ${\sigma}_{\rm nm}$ characterize the damping due to the formation of the flux-line lattice (FLL) in the SC state and of the nuclear magnetic dipolar contribution, respectively. During the analysis ${\sigma}_{\rm nm}$ was assumed to be constant over the entire temperature range and was fixed to the value obtained above $T_{\rm c}$ where only nuclear magnetic moments contribute to the muon depolarization rate ${\sigma}$.

\begin{figure}[t!]
\includegraphics[width=1.0\linewidth]{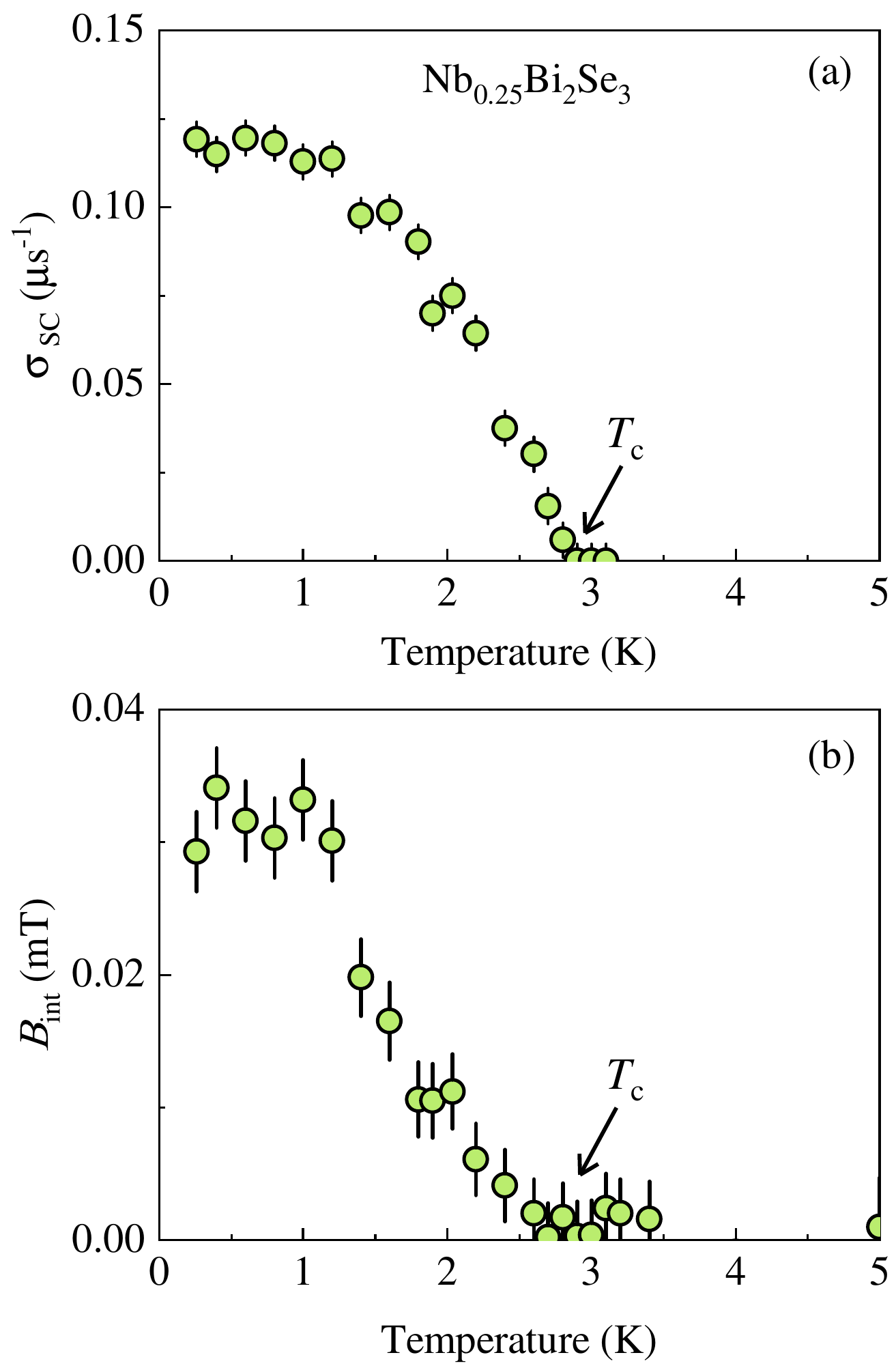}
\caption{(Color online) \textbf{Superconducting muon spin depolarization rate ${\sigma}_{\rm sc}$ and the field shift for Nb$_ {0.25}$Bi$_{2}$Se$_{3}$.}  (a) Temperature dependence of the  superconducting muon spin depolarization rate
${\sigma}_{\rm sc}$ measured in an applied magnetic field of ${\mu}_{\rm 0}H = 0.01$~T for Nb$_ {0.25}$Bi$_{2}$Se$_{3}$. The error bars represent the s.d. of the fit parameters. (b) Temperature dependence of the difference between the internal field ${\mu}_{\rm 0}$$H_{\rm SC}$ measured in the SC state and the one measured in the normal state ${\mu}_{\rm 0}$$H_{\rm NS}$ at $T$ = 5~K for Nb$_ {0.25}$Bi$_{2}$Se$_{3}$.}
\label{fig7}
\end{figure}

In Fig.~3a, ${\sigma}_{\rm sc}$ is plotted as a function of temperature for Nb$_ {0.25}$Bi$_{2}$Se$_{3}$ at ${\mu}_{\rm 0}H=0.01$~T. Below $T_{\rm c}$ the relaxation rate ${\sigma}_{\rm sc}$ starts to increase from zero due to the formation of the FLL and shows the saturation towards the low temperatures. We show in the following paragraph that the observed temperature dependence of ${\sigma}_{{\rm sc}}$, which reflects the topology of the SC gap, is consistent with the presence of the single SC gap on the Fermi surface of Nb$_ {0.25}$Bi$_{2}$Se$_{3}$. It is interesting to note that below $T_{\rm c}$, we observed a paramagnetic shift of the internal magnetic field $B_{int}$, sensed by the muons, instead of the expected diamagnetic shift imposed by the SC state. This is evident in Fig. 3b, where we plot the difference between the internal field $B_{int,SC}$ measured in the SC state and one $B_{int,NS}$ measured in the normal state $T$ = 5 K.

To proceed with a quantitative analysis, we note that the temperature dependence of the London magnetic penetration depth ${\lambda}(T)$ is related to the muon spin depolarization rate ${\sigma}_{\rm sc}(T)$ in the presence of a perfect triangular vortex lattice by the equation:\cite{Brandt}
\begin{equation}
\frac{\sigma_{sc}^2(T)}{\gamma_\mu^2}=0.00371\frac{\Phi_0^2}{\lambda^4(T)},
\end{equation}
where ${\Phi}_{\rm 0}=2.068 {\times} 10^{-15}$~Wb is the magnetic-flux quantum. Equation (2) is only valid when the separation between the vortices is smaller than ${\lambda}$. In this case, according to the London model, ${\sigma}_{\rm sc}$ is field independent\cite{Brandt}.

To investigate the superconducting gap structure of Nb$_ {0.25}$Bi$_{2}$Se$_{3}$, we analyzed the $T$-dependence of the magnetic penetration depth, ${\lambda}(T)$, which is directly associated with the superconducting gap.  ${\lambda}$($T$) can be described within the local (London) approximation (${\lambda}$ ${\gg}$ ${\xi}$) by the following expression:\cite{Bastian,Tinkham}
\begin{equation}
\frac{\lambda^{-2}(T,\Delta_{0,i})}{\lambda^{-2}(0,\Delta_{0,i})}=
1+\frac{1}{\pi}\int_{0}^{2\pi}\int_{\Delta(_{T,\varphi})}^{\infty}(\frac{\partial f}{\partial E})\frac{EdEd\varphi}{\sqrt{E^2-\Delta_i(T,\varphi)^2}},
\end{equation}
where $f=[1+\exp(E/k_{\rm B}T)]^{-1}$ is the Fermi function, ${\varphi}$ is the angle along the Fermi surface, and ${\Delta}_{i}(T,{\varphi})={\Delta}_{0,i}{\Gamma}(T/T_{\rm c})g({\varphi}$) (${\Delta}_{0,i}$ is the maximum gap value at $T=0$).
The temperature dependence of the gap is approximated by the expression ${\Gamma}(T/T_{\rm c})=\tanh{\{}1.82[1.018(T_{\rm c}/T-1)]^{0.51}{\}}$,\cite{carrington} while $g({\varphi}$) describes the angular dependence of the gap and it is replaced by 1 for both an $s$-wave gap, [1+acos(4${\varphi}$))/(1+a)] for an anisotropic $s$-wave gap and ${\mid}\cos(2{\varphi}){\mid}$ for a $d$-wave gap.\cite{Fang}

As seen in Fig~4a, the experimentally obtained ${\lambda}$($T$) dependence is best described by a momentum independent $s$-wave model with a gap value of $\Delta$ = 0.49(1)~meV and $T_{\rm c}$ = 2.86(2)~K. We note that $T_{\rm c}$ obtained from $\mu$SR is very close to those derived from susceptibility and specific heat measurements (see Fig. 1b). An anisotropic $s$-wave model, which would be consistent with the recently proposed fully gapped $\Delta_{4y}$ case, where the superconducting gap has deep minima along the $k_{x}$ direction, describes the experimental data but only with small angular variation of the gap.
We note that the ($p_{\rm x}$ + i$p_{\rm y}$) pairing symmetry is also characterised by the full gap in 2D systems and would also give saturated behaviour at low temperatures. However, the possibility of $p_{\rm x}$ + i$p_{\rm y}$ pairing is excluded by the absence of TRS breaking state.
Also $d$-wave gap symmetry was tested, but was found to be inconsistent with the data (for example, see the dashed line in Fig~4a). In particular, it is difficult to account for the very weak temperature dependence of ${\lambda}$($T$) at low-$T$ within such models. We also tested the power law (1-(T/$T_{c}$)$^{2}$ which has been proposed theoretically \cite{Hirshfeld} for the superfluid density of dirty $d$-wave superconductors and found it to be inconsistent with the data (see Fig. 4a).
This leaves a nodeless or fully gapped state as the most plausible bulk SC pairing state in the bulk of 
Nb$_{0.25}$Bi$_{2}$Se$_{3}$. It should be noted, though, that saturation of the muon depolarization rate can arise under certain conditions even in nodal superconductors \cite{Sonier,KhasanovPRB,Luetkens,SonierJPSJ,Harshman} and that in such cases  extensions beyond Eq. 2 may be required to extract the superfluid density. In this particular case, fully gapped state is substantiated by the field dependent measurements. Figure 4b shows the obtained field dependence of ${\sigma}_{sc}$ at 250 mK. Each point was obtained by field cooling the sample from above $T_{\rm c}$ to 250 mK. As expected from the London model one observes the maximum at 10 mT. Above 10 mT, ${\sigma}_{sc}$ decreases with increasing magnetic field. This appears consistent with a behaviour expected for $s$-wave superconductor for an ideal triangular vortex lattice \cite{Brandt}. Taking into account the critical field of 0.9 T, the theoretical $s$-wave behaviour has been calculated according to the Brandt formula \cite{Brandt}. It is shown by the solid black line in Fig. 4b, which agrees well with the experimental data. The model with nodes in the gap (dashed blue line), taking into account nonlocal effects \cite{Luetkens}, was also tested and found to be inconsistent with the data. Thus, combination of temperature and field dependent measurements of ${\sigma}_{sc}$ provide the strong evidence for the fully gapped state. The ratio of the SC gap to $T_{\rm c}$ was estimated to be (2$\Delta/k_{\rm B}T_{\rm c}$) ${\simeq}$ 3.95, which is consistent with the strong coupling limit BCS expectation \cite{GuguchiaMoTe2}. However, a similar ratio can also be expected for Bose Einstein Condensation (BEC)-like picture. It is important to note that the ratio $\Delta/k_{\rm B}T_{\rm c}$ does not effectively distinguish between BCS or BEC condensation. Having in mind 100 ${\%}$ SC volume fraction, the effective penetration depth, $\lambda_{eff}$, at zero temperature is found to be 930(10)~nm. 
Such a high value of $\lambda_{eff}$ was also estimated previously for other sister compounds namely Cu and Sr doped Bi$_2$Si$_3$ \cite{Krieger,Leng,Neha}. In view of the short coherence length and relatively large mean free path $l$ ${\sim}$  50 nm \cite{Smylie3}, we can assume that Nb$_ {0.25}$Bi$_{2}$Se$_{3}$ lies close to the clean limit. With this assumption, we obtain the ground-state value of the carrier density $n_{s}/(m^{*}/m_{e}$) ${\simeq}$ 0.25 ${\times}$ 10$^{26}$ m$^{-3}$. This density is by factor of ${\sim}$ 7 smaller than the one 1.67 ${\times}$ 10$^{26}$ m$^{-3}$ obtained for the unconventional Weyl-superconductor $T_{d}$-MoTe$_{2}$ \cite{GuguchiaMoTe2} with the similar $T_{\rm c}$. This fact indicates the relatively high $T_{\rm c}$ for a small number of carriers and $T_{\rm c}$/$\lambda_{eff}^{-2}$ ratio is comparable to those of high-temperature unconventional superconductors \cite{Uemura1,Uemura2}.

\begin{figure}[htb!]
\includegraphics[width=1.0\linewidth]{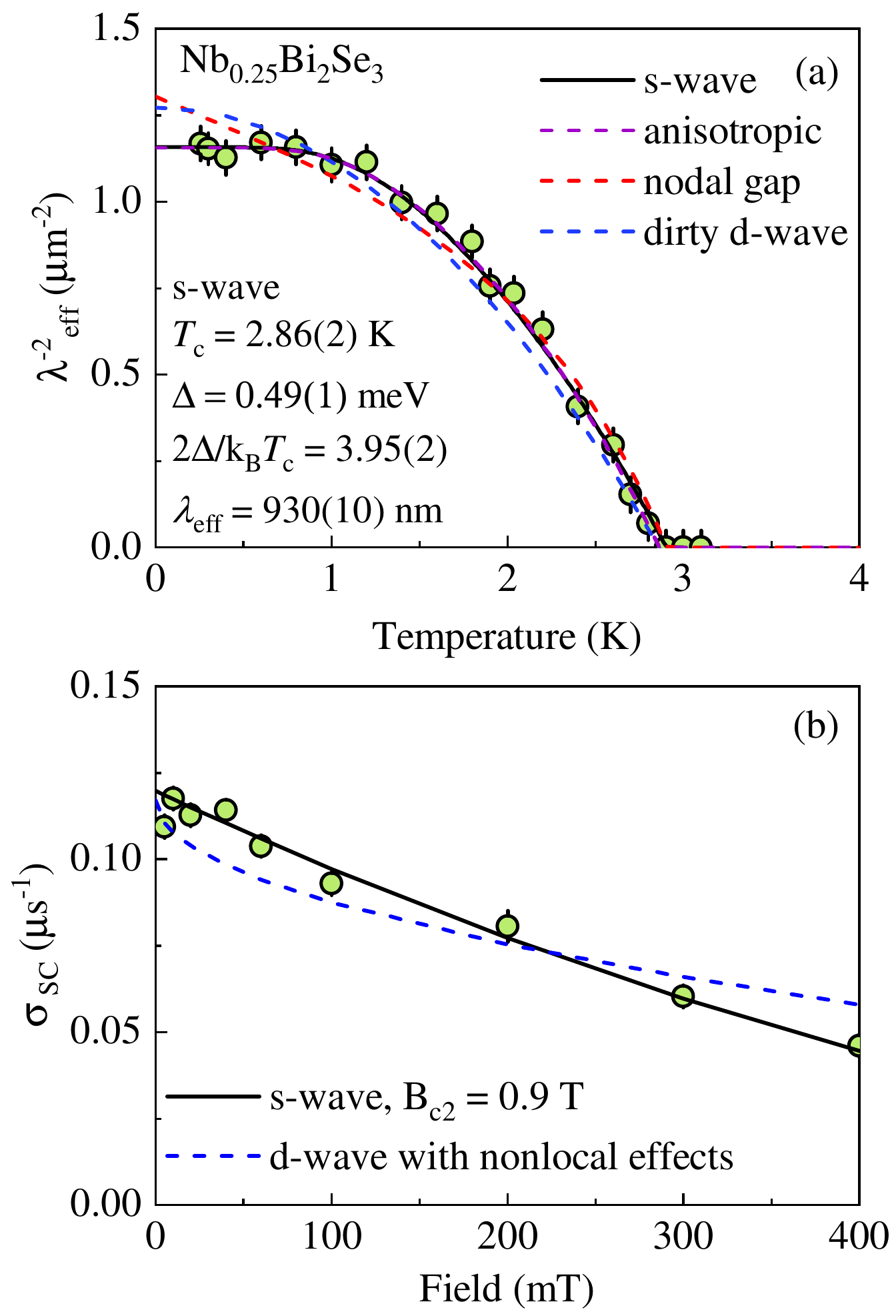}
\caption{(Color online) \textbf{Temperature and field evolution of ${\lambda}^{-2}$ for Nb$_ {0.25}$Bi$_{2}$Se$_{3}$.} (a) The temperature dependence of ${\lambda}^{-2}$ for Nb$_ {0.25}$Bi$_{2}$Se$_{3}$, measured in an applied field of ${\mu}_{\rm 0}H=0.01$~T. The solid black line corresponds to a single gap BCS $s$-wave model. The dashed purple line corresponds to anisotropic $s$-wave model. The dashed red line corresponds to clean limit $d$-wave model. The dashed blue line corresponds to a dirty limit $d$-wave model (a power law (1-(T/$T_{c}$)$^{n}$ with $n$ = 2). The error bars are calculated as the s.e.m. (b) Field dependence of the superconducting muon spin depolarization rate 
${\sigma}_{\rm sc}$.}
\label{fig7}
\end{figure}

In order to support the conclusion derived from $\mu$SR investigation, we have performed an analysis of the superconducting part of specific heat. The superconducting part of the specific heat $C_{el}$ (= $C_{S}$-$C_{N}$  ) is related to the the entropy $S_{sc}$ in the superconducting state by the relation $C_{el}  = T \left(\frac{\delta S_{sc}}{\delta T}\right)$. Now, within the formulation of BCS theory, $S_{sc}$ can be written as\cite{Nakajima, Chen} $S_{sc}  = - \frac{3\gamma_n}{\kappa_B\pi^3}\int_0^{2\pi}\int_0^{\infty}\left[(1-f) ln(1-f)+f lnf\right] d\varepsilon d\phi$, where $\gamma_n$ is the normal state Sommerfeld coefficient, $f$ is the Fermi function with energy $E$ = $\sqrt{\varepsilon^2+\Delta^2(\phi,T)}$. Therefore, applying the same methodology as adapted for analyzing $\lambda(T)$, we can analyze the specific heat data. Fig~5 depicts that, over the range of measured temperature, the superconducting contribution to the specific heat can also be modeled by a fully gapped $s$-wave picture, yielding $\Delta/k_{\rm B}T_{\rm c}$ = 1.9, consistent with the $\mu$SR measurements. However this result has to be taken cautiously since we could not access the low temperature region of the specific heat.
 
\begin{figure}[htb!]
\includegraphics[width=0.9\linewidth]{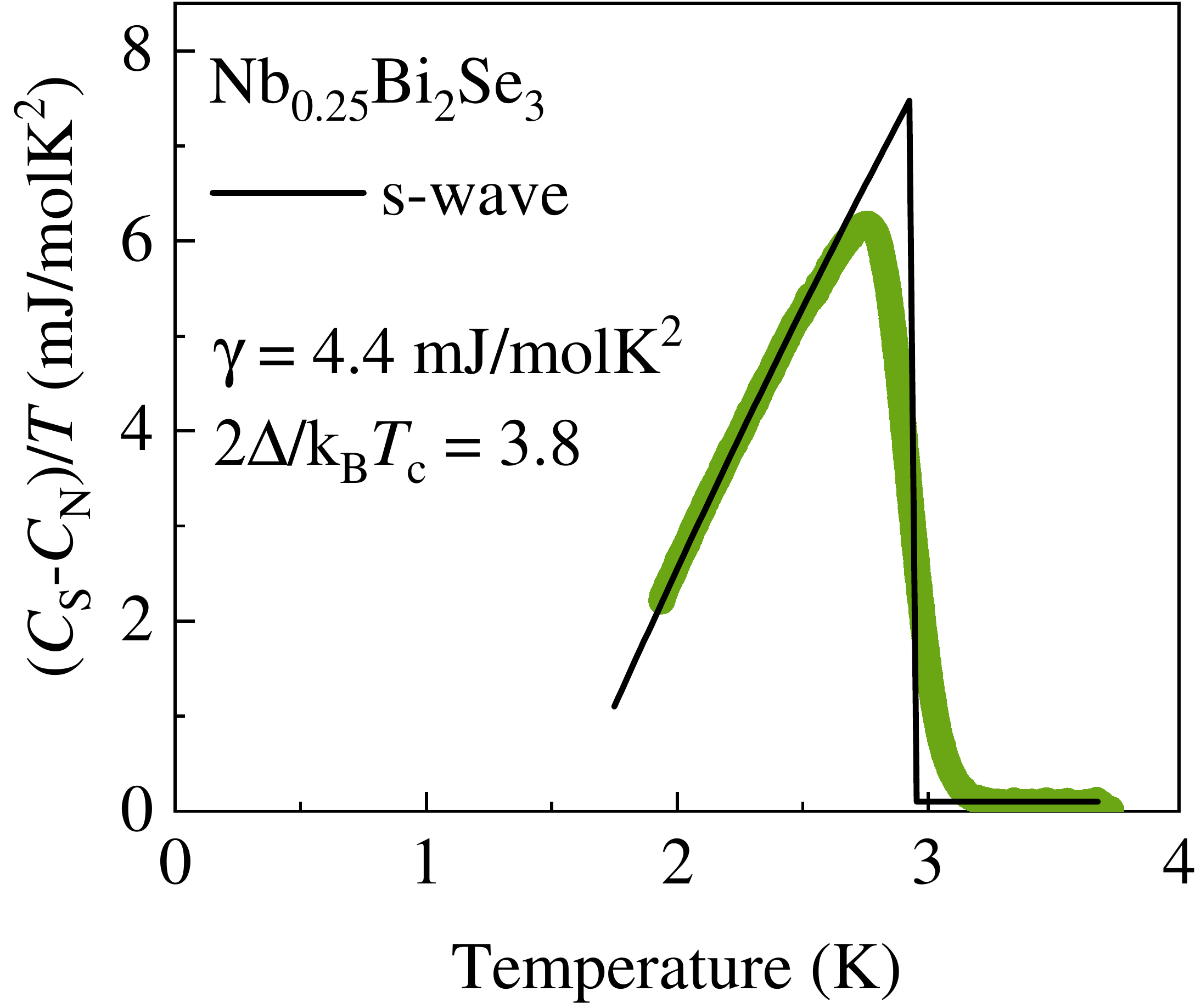}
\caption{(Color online) \textbf{Temperature evolution of specific heat for Nb$_ {0.25}$Bi$_{2}$Se$_{3}$.} 
Temperature evolution of the electronic part of specific heat fitted with single $s$-wave gap model (solid line).}
\label{fig7}
\end{figure}
 
\section{DISCUSSION}

One of the findings of this paper is the absence of magnetism and time-reversal symmetry breaking state in the bulk of Nb$_ {0.25}$Bi$_{2}$Se$_{3}$ below $T_{\rm c}$, classifying this system as a non-magnetic and time-reversal invariant superconductor. This is in contrast with a previous report \cite{Qiu} which suggested a TRS breaking state with triplet chiral $p$-wave pairing. Such conclusion was made based on the observation of anomalous Hall resistivity below $T_{\rm c}$  and taking into account some theoretical aspects of impurity scattering in $p$-wave superconductors. The discrepancy between our results and the previous report might be related to the fact that the Hall response may contain surface effects, while with ${\mu}$SR we are probing deep in the bulk. In addition, the anomalous Hall response within the SC state may originate not from magnetism, but rather from an intrinsic topological band structure and Berry curvature effects. 

In addition to the TRS invariant SC state, another interesting observation is the fact that the $T_{\rm c}$/$\lambda_{eff}^{-2}$ ratio for Nb$_ {0.25}$Bi$_{2}$Se$_{3}$ is comparable to those of high-temperature unconventional superconductors. 
Within the picture of Bose Einstein Condensation (BEC) to BCS crossover \cite{Uemura1,Uemura2}, systems exhibiting small $T_{\rm c}$/$\lambda_{eff}^{-2}$ ${\sim}$ 0.00025-0.015 are considered to be in the BCS-like side, while the large value of $T_{\rm c}$/$\lambda_{eff}^{-2}$ ${\sim}$ 1-20 and the linear relationship between $T_{\rm c}$ and $\lambda_{eff}^{-2}$ is expected only in the BEC-like side and is considered a hallmark feature of unconventional superconductivity. This approach has in the past been used for the characterization of BCS-like, so-called conventional and BEC-like, so-called unconventional superconductors. The here present results on Nb$_ {0.25}$Bi$_{2}$Se$_{3}$ demonstrate the ratio $T_{\rm c}$/$\lambda_{eff}^{-2}$ ${\sim}$ 2.5, which is close to the one $T_{\rm c}$/$\lambda_{eff}^{-2}$ ${\sim}$ 4 for hole-doped cuprates. This result provides strong evidence for an unconventional pairing mechanism in Nb$_ {0.25}$Bi$_{2}$Se$_{3}$, a system which exhibits the topologically non-trivial electronic structure. In addition, there are few other signatures of unconventional superconductivity in  Nb$_ {0.25}$Bi$_{2}$Se$_{3}$. Namely, (1) Similar to other unconventional superconductors, multiple-orbit nature of the electronic state \cite{Lawson} was observed in Nb$_ {0.25}$Bi$_{2}$Se$_{3}$, which might point to Fermi surface nesting as a possible superconducting mechanism. (2) The emergence of a pronounced two-fold in-plane anisotropy of all superconducting properties and power-law temperature dependence of the penetration depth at the surface \cite{Smylie1,Smylie2}. (3) Unusual paramagnetic shift (this work). (4) A substantial increase in electron scattering by disorder is required to suppress $T_{\rm c}$ in Nb$_ {0.25}$Bi$_{2}$Se$_{3}$, far larger than anticipated via conventional theory \cite{Smylie2}. (5) Nb$_ {0.25}$Bi$_{2}$Se$_{3}$ manifests a very unusual upper critical field behaviour signaling an unconventional nature of superconducting state in this compound \cite{Kobayashi}. Similar to many unconventional superconductors, including the topological Weyl-superconductor $T_{d}$-MoTe$_{2}$  \cite{GuguchiaMoTe2}, we observed fully gapped superconducting state in the bulk of Nb$_ {0.25}$Bi$_{2}$Se$_{3}$. The combination of unconventional superconductivity, fully gapped SC and TRS invariant state in the bulk of Nb$_ {0.25}$Bi$_{2}$Se$_{3}$ can be consistent with the particular class of topological superconducting phase belonging to the same symmetry as the Balian-Werthamer (BW) state \cite{Balian}, i.e., similar to the B-phase of $^{3}$He. Note that we observed only  one SC gap in this multi band system Nb$_ {0.25}$Bi$_{2}$Se$_{3}$.  ARPES and Hall effect measurements revealed the n-type conductivity in the normal state of Nb$_ {0.25}$Bi$_{2}$Se$_{3}$ and suggested that a single type of electron carriers are dominant with the normal state carrier density \cite{Qiu,Lawson} $n_{e}$ ${\simeq}$ 1.5 ${\times}$ 10$^{26}$ m$^{-3}$. The Hall carrier density is generally an order of magnitude larger than the carrier density given by the ellipsoidal Fermi surfaces. The carrier density of the ellipsoidal Fermi surfaces was reported \cite{Lawson} $n_{e}$ ${\simeq}$ 2.6 ${\times}$ 10$^{25}$ m$^{-3}$ for Nb$_ {0.25}$Bi$_{2}$Se$_{3}$. It is interesting that the SC carrier density $n_{s}$ (using $m^{*}/m_{e}$ = 0.26, maximum value from Ref.~\cite{Lawson}) ${\simeq}$ 0.7 ${\times}$ 10$^{25}$ m$^{-3}$, which we estimated for Nb$_ {0.25}$Bi$_{2}$Se$_{3}$ is very close to the normal state electron density, indicating that the superconductivity stems mostly from electron carriers. Thus, single gap superconductivity in Nb$_ {0.25}$Bi$_{2}$Se$_{3}$ may be explained by superconducting gap occurring only on the electron-like Fermi surface.

Regarding the gap symmetry in Nb$_{0.25}$Bi$_{2}$Se$_{3}$, using the tunnel-diode oscillator(TDO) technique Smylie $et.~al.$\cite{Smylie1,Smylie2} showed that change in London penetration depth $\Delta\lambda(T)$ follows a power law 
$\Delta\lambda(T)\sim~T^2$  dependence indicating the presence of symmetry-protected point nodes. However, it has to be noted that TDO is a surface sensitive technique whereas $\mu$SR is a bulk probe. More precisely, the response of a superconductor in the Meissner state to an AC electromagnetic field with frequency far below the gap frequency is described by the temperature dependent anisotropic London penetration depth \cite{Prozorov}. Therefore, in a TDO experiment, only material in the near-surface area to the order of a few ${\lambda}$ is probed while in a $\mu$SR experiment the whole sample volume is probed. The $\mu$SR experiments presented here yield an upper limit of lambda of $\sim$ 900 nm, while previous determinations based on measurements of the lower critical field yielded $\sim$ 250 nm for the in-plane penetration depth \cite{Smylie1}. Typically, TDO measurements are interpreted as representing the bulk behavior of the superconductor. However, recent theoretical analysis by Wu $et~al$ \cite{Wu}, suggests that the formation of a surface state at the surface of a topological superconductor causes an effective penetration depth describing the electromagnetic fields near the surface which contains a correction term with power-law temperature dependence. The authors considered the fully gapped SC state in the bulk. In addition to the bulk-dominated London response, they identified an additional $\Delta\lambda(T)\sim~T^3$ power-law-in-temperature contribution from the surface, valid in the low-temperature limit, and it was argued that the power-law temperature dependence of the penetration depth can be one indicator of topological superconductivity. Since TDO probes the near-surface electromagnetic fields it will sense this correction and non-exponential temperature dependence while this will go completely unnoticed in $\mu$SR or specific heat.  Currently, it is not clear why the large in-plane anisotropy, which is an established bulk effect in Nb$_{0.25}$Bi$_{2}$Se$_{3}$, is not reflected in the bulk SC gap structure. Future work will be needed to address this important issue.


Finally, the observation of the paramagnetic shift in the SC state of Nb$_ {0.25}$Bi$_{2}$Se$_{3}$, instead of the expected diamagnetic shift imposed by the SC state, deserves some attention. In general, the paramagnetic shift can have few possible causes: field induced magnetism \cite{Khasanov,GuguchiaPRB}, vortex disorder \cite{Sonier2011}, odd-frequency superconducting state \cite{Salman}, and by the suppression of the negative knight shift below $T_{\rm c}$ due to singlet pairing. Considering the absence of any trace of magnetism in Nb$_{0.25}$Bi$_{2}$Se$_{3}$ and the symmetric line-shape from TF-${\mu}$SR, magnetism and disorder effects can be excluded as being the origin of the observed paramagnetic shift. The shape of the ${\sigma}_{\rm sc}(T)$ is also not consistent with odd-frequency superconducting pairing. Additional experiments are needed to explore the origin of such an unconventional paramagnetic shift in 
Nb$_ {0.25}$Bi$_{2}$Se$_{3}$.\\

\section{CONCLUSIONS}

In conclusion, we provide a microscopic investigation of the superconductivity in the layered topological superconductor candidate Nb$_{0.25}$Bi$_{2}$Se$_{3}$ with a bulk probe. Specifically, the zero-temperature magnetic penetration depth ${\lambda}_{eff}\left(0\right)$ and the temperature as well as the field dependence of ${\lambda_{eff}^{-2}}$ were studied by means of ${\mu}$SR experiments. The superfluid density can be described in a scenario of a complete gap. Though, future work will explore the relation between the measured muon depolarization rate and the penetration depth in anisotropic and possibly inhomogenous systems. Interestingly, $T_{\rm c}$/$\lambda_{eff}^{-2}$ ratio is comparable to those of high-temperature unconventional superconductors, pointing to the unconventional nature of superconductivity in Nb$_{0.25}$Bi$_{2}$Se$_{3}$. Furthermore, ${\mu}$SR, which is extremely sensitive magnetic probe, does not show evidence of any spontaneous magnetic fields that would be expected for a TRS breaking state  in the bulk of Nb$_ {0.25}$Bi$_{2}$Se$_{3}$. Our results classify Nb$_ {0.25}$Bi$_{2}$Se$_{3}$ as unconventional time-reversal-invariant and fully gapped bulk superconductor.\\



\section{Acknowledgments}~
The work was performed at the Swiss Muon Source (S${\mu}$S) Paul Scherrer Insitute, Villigen, Switzerland.
Work in Okayama is supported by JSPS KAKENHI Grant Number 18K03540 and 19H01852. 
M.Z.H. acknowledges visiting scientist support from IQIM at the California Institute of Technology.
KW acknowledges funding from the SNSF through a postdoc mobility fellowship. Specific heat measurements were supported by the U.S. Department of Energy, Office of Science, Basic Energy Sciences, Materials Sciences and Engineering Division.\\



\end{document}